\documentclass[aps,prd,showpacs,reprint,notitlepage,superscriptaddress]{revtex4-1}
\usepackage[colorlinks=true,allcolors=blue]{hyperref}
\usepackage{amsmath,amsthm,amssymb,amsopn}
\usepackage{slashed}
\usepackage{bm,bbm}
\usepackage[dvips,final]{graphicx}
\usepackage{url}
\usepackage{subfigure}
\usepackage{textcomp}
\usepackage{dcolumn}
\usepackage[usenames]{color,xcolor}
\usepackage{ulem}
\usepackage{cancel}
\usepackage{epsfig}
\usepackage{braket}
\usepackage[title]{appendix}
\usepackage{mathptmx}
\DeclareMathAlphabet{\mathcal}{OMS}{cmsy}{m}{n}

\def\comment#1{}

\def\l{\left}
\def\r{\right}
\def\beq{\begin{equation}}
\def\eeq{\end{equation}}
\def\bea{\begin{eqnarray}}
\def\eea{\end{eqnarray}}

\def\ep{\epsilon\!\!\!/}
\def\transpose{^\mathsf{T}}

\def\red#1{\textcolor{red}{#1}}

\input epsf.tex

\def\ep{\epsilon\!\!\!/}
\def\comment#1{}


\begin{document}

\title{Distinguishing Dirac from Majorana neutrinos in a microwave cavity}

\author{Mehdi Abdi}
\email[]{mehabdi@cc.iut.ac.ir}
\affiliation{Department of Physics, Isfahan University of Technology, Isfahan 84156-83111, Iran}

\author{Roohollah Mohammadi}
\email[]{rmohammadi@ipm.ir}
\affiliation{Iranian National Museum of Science and Technology (INMOST), PO Box 11369-14611, Tehran, Iran}
\affiliation{School of Astronomy, Institute for Research in Fundamental Sciences (IPM), PO Box 19395-5531, Tehran, Iran}

\author{She-Sheng Xue}
\email[]{xue@icra.it ; shesheng.xue@gmail.com}
\affiliation{ICRANet Piazzale della Repubblica, 10 65122 Pescara, Italy}
\affiliation{Physics Department, University of Rome La Sapienza, P.le Aldo Moro 5, I00185 Rome, Italy}

\author{Moslem Zarei}
\email[]{m.zarei@cc.iut.ac.ir}
\affiliation{Department of Physics, Isfahan University of Technology, Isfahan 84156-83111, Iran}

\affiliation{School of Astronomy, Institute for Research in Fundamental Sciences (IPM), PO Box 19395-5531, Tehran, Iran}

\affiliation{ICRANet-Isfahan, Isfahan University of Technology, 84156-83111, Iran}

\date{\today}

\begin{abstract}
We propose a novel scheme for distinguishing between the Dirac and Majorana nature of neutrinos via interaction of a neutrino beam with microwave photons inside a cavity.
We study the effective photon-photon polarization exchange induced by the photon-neutrino scattering.
The quantum field theoretical studies of such effective picture are presented for both Dirac and Majorana neutrinos.
Our phenomenological analyses show that the difference between Dirac and Majorana neutrinos can manifest itself in scattering rate of the photons.
To enhance the effect a cavity scheme is employed.
An experimental setup based on microwave cavities is then designed and simulated by finite element method to measure the scattering rate. Our results suggest that an experiment based on the current state-of-the-art technology will be able to probe the difference in about one year. However, it can be done in a few days by enhancing the neutrino beam flux or implementing with the near future equipments.
Therefore, our work provides the possibility 
for solving the long lasting puzzle of Dirac or Majorana nature of neutrinos.
\end{abstract}


\maketitle

\section{Introduction}
Since neutrinos were introduced by Pauli in 1930 to explain the energy spectrum of electrons in beta decays, they have always been extremely peculiar in the Standard Model (SM) for elementary particle physics. Their properties of charge neutrality, near masslessness, flavor mixing, oscillation, the Dirac or Majorana nature, and in particular the parity-violating gauge-coupling have been at the center of theoretically conceptual elaborations and developments for almost a century. On the other hand, to reveal neutrino nature and properties, sophisticated experiments, observations and data analyses have been made and ongoing for many decades. The progress and the results have played an essential role in understanding the neutrino physics in the cosmology, astrophysics, nuclear and elementary particle physics.

In the past years, the several neutrino experiments based on the reactor, solar and atmospheric neutrino sources have shown and confirmed the evidence of neutrino flavor oscillations \cite{osc1}. This fact implies an important point that neutrinos are not exactly massless, although they chirally couple to $W$ and $Z$ gauge bosons in the SM. Therefore the neutrinos cannot be exactly described as two-component Weyl fermions. They should be either four-component Dirac or Majorana neutrinos \cite{majorana}.

The question of whether the neutrino is Dirac or Majorana type is the most important and longstanding issue in not only the SM theory, but also experiment. As will be discussed in this article, the probing neutrino electromagnetic interactions are the best way to distinguish between Dirac and Majorana neutrinos, since Dirac and Majorana neutrinos have different electromagnetic properties \cite{nieves}.
The photon-neutrino couplings have been extensively studied in the literature (see for example \cite{Broggini:2012df}), they are extremely small due to the electroweak coupling at the one-loop level. As a consequence, it is really hard to measure photon-neutrino couplings in the manner of traditional nuclear and particle experiments.
The well-known example is the experiment of the neutrino-less double beta decay \cite{dbd}.

Since the laser and microwave technologies have been greatly improved and sophisticated in recent years, a lot of attention has been driven to study the neutrino-photon coupling based on the interactions between neutrinos and laser beams. As examples, Ref.~\cite{titov} has analyzed the emission of $\nu\bar\nu$ pairs off electrons in a polarized ultra-intense electromagnetic (e.g., laser) wave field. In Ref.~\cite{tinsley}, by using neutrinos interacting with photons from an intense laser field, the electron-positron production rate has been studied. This field seems very promising and a new window, in addition to the traditional experiments in nuclear and particle physics.

It is shown \cite{Mohammadi, Mohammadi:2013ksa} that linearly polarized photons acquire their circular polarization via forward scatterings with neutrinos.
This turns out to be crucial for investigating the properties of Majorana neutrinos, which have no electric and magnetic dipole moments due to CPT invariance \cite{palbook}.
In Ref.~\cite{Mohammadi:2013ksa}, it is pointed out that only left-handed neutrino coupling to $W$-boson is the reason for photons acquiring circular polarization, through the photon-neutrino interacting vertex, see Fig.~\ref{diagram}. The acquired circular polarization is described by the Stokes parameter $\Delta V$, and the $\Delta V$-amplitude of linearly polarized laser photons forward scattering off Dirac or Majorana neutrino are obtained.
The former ``$D$'' is twice smaller than the latter ``$M$'', i.e., $\Delta V^D=\Delta V^M/2$, due to their difference in the degree of freedoms.
In contrast with the extreme smallness of photon-neutrino interacting cross section, the forward scattering amplitudes of photon circulation polarization are hopefully detectable by using laser and microwave technologies.
Moreover, by considering available neutrino and ultra-intense laser beams, it is shown that the circular polarization amplitude $\Delta V^{D,M}$ of a scattered laser beam off a Dirac or Majorana neutrino beam $\Delta V^{D,M}\propto\lambda\times\Delta t$, where $\lambda$ is the photon wavelengths and $\Delta t$ is the time duration of photon-neutrino interaction. It is thus concluded that the increasing $\Delta t$ by the back and forth interactions should open a window of detecting $\Delta V^{D,M}$, so as to distinguish between Dirac and Majorana neutrinos.

In this article, we examine the possibility of detecting polarization scattering of the photons induced by the incidence of a neutrino beam. The scattering is described in the term of an effective interaction Hamiltonian where two linearly polarized microwave modes couple to each other.
The coupling, in turn, is induced by the photon-neutrino scattering amplitude in the manners of both one-loop calculations and general derivation of the covariant form based on the CPT symmetries.
Apart from small and unimportant corrections to the kinetic terms of linear polarized photons, the effective Hamiltonian presents the coupling of two orthogonal polarization modes of the photons. The coupling is inversely proportional to the frequency of the photons, favoring low frequency domains.
Therefore, in order to measure the rate of polarization exchange in the cavity we design an experiment based on microwave cavities. The microwave cavities are among those with the highest quality factors. Their quality factor can even reach $10^{10}$~\cite{Romanenko2014}, which means a photon bounces back and forth ten trillion times from the cavity walls before leaving it.
This provides higher coherency in the experiment, and thus, a better resolution in detecting the coupling rate.
We support our theory by finite element simulation of the cavity modes and discuss that by employing three different cavity modes the measurement can be performed.
A transmon qubit, state-of-the-art superconducting qubit with long coherence time, is proposed for probing and detecting the scattered photons.

The paper is organized as follows: In the next section we derive the photon-neutrino scattering amplitude for both Dirac and Majorana neutrinos.
In Sec.~\ref{sec:hamiltonian} an effective Hamiltonian is for the neutrino-induced photon-photon scattering.
In Sec.~\ref{sec:cavity} the Hamiltonian is used to analyze the scattering rate and possibility of resolving the two neutrino candidates and Sec.~\ref{sec:implement} is devoted to discussion about the designed experiment.
The paper ends with outlook and concluding remarks in Sec.~\ref{sec:conclusion}.

\section{Invariant photon-neutrino scattering amplitude \label{sec:amplitude}}
Here, we find a general amplitude for the photon-neutrino scattering process
$\gamma(p,s)+\nu(q,r)\rightarrow \gamma(p',s')+\nu(q',r')$
using the method developed in \cite{Lifshitz,Prange,Latimer:2016kdg,zarei}. 
The Lorentz invariant amplitude of such a process can generally be written in the form \cite{Latimer:2016kdg,zarei}
\beq
T_{fi}(ps,qr,p's',q'r')=\epsilon^s_{\mu}(p)\epsilon^{s'}_{\nu}(p')\bar{u}_{r'}(q')F^{\mu\nu}u_r(q)~,
\eeq
where $u_r(q)$ is the neutrino spinor with spin index $r=1,2$ and $\epsilon^s_{\mu}(p)$ is the photon polarization vectors.
The general rank-two tensor $F^{\mu\nu}$ is constructed using the set of space-like basis vectors $\hat{e}^{(1)}$ and $\hat{e}^{(2)}$, satisfying the orthogonality condition $\hat{e}^{(1)}\cdot \hat{e}^{(2)}=0$. In order to construct these two vectors we use the kinematics variables $Q^{\lambda}=(q^{\lambda}+q'^{\lambda})-\frac{P^{\lambda}}{P^{2}}\:(q+q')\cdot P$, $N^{\lambda}=\epsilon^{\lambda\mu\nu\rho}Q_{\mu}t_{\nu}P_{\rho}$, $P^{\lambda}=p^{\lambda}+p'^{\lambda}$ and  $t^{\lambda}=q^{\lambda}-q'^{\lambda}=p'^{\lambda}-p^{\lambda}$. Then the normalized basis vectors are given by $\hat{e}^{(1)\lambda}=N^{\lambda}/\sqrt{-N^2}$ and $\hat{e}^{(2)\lambda}=Q^{\lambda}/\sqrt{-Q^2}$ (see Ref.~\cite{zarei} for more details).
In the following, we derive general scattering amplitude for Dirac and Majorana neutrinos.

\subsection{Dirac neutrino}
We assume a Dirac spinor for neutrino and consider two special cases: (i) the amplitude is odd under parity, and (ii) the amplitude is even under parity. As we pointed above, for both cases (i) and (ii) the amplitude must be even under CPT.
We first consider the case (i) and impose the odd-parity condition on the amplitude. According to the CPT invariance, we further impose T-even and C-odd condition. As a result, the amplitude will be even
under CP and CPT. The most general form of $F_D^{\mu\nu}$ that is odd under P and even under T is determined up to four real coefficients $f_i$
\bea
F_{D}^{\mu\nu} &&=f_1\gamma^{5}\slashed{P}\,\hat{e}^{(1)\mu}\hat{e}^{(1)\nu}+f_2\gamma^{5}\slashed{P}\,\hat{e}^{(2)\mu}\hat{e}^{(2) \nu} \nonumber\\
&&+\left(f_3+f_4\slashed{P}\right)\left[\hat{e}^{(1) \mu}\hat{e}^{(2) \nu}+\hat{e}^{(2) \mu}\hat{e}^{(1) \nu}\right]~,
\label{amplitude1}
\eea
and the C-even condition gives no more constraint on the coefficients \cite{Latimer:2016kdg,zarei}. The conditions of C-even and crossing symmetry should be imposed on photon fields, 
\beq
F_{D}^{\mu\nu}(q',p';q,p)=-C\left[F_{D}^{\mu\nu}(-q,-p;-q',-p')\right]\transpose C^{-1}~.
\label{conpc}
\eeq
The amplitude \eqref{amplitude1} satisfies the condition (\ref{conpc}) and gives non-vanishing circular polarization for CMB radiation \cite{zarei}. For the case (i), it is also interesting to
examine the condition of C-even and T-odd in which the amplitude is PT-even.
For this case, we find \cite{zarei}
\bea
F_{D}^{\mu\nu} &&= f_1\gamma^{5}\,\hat{e}^{(1)\mu}\hat{e}^{(1)\nu}+f_2\gamma^{5}\,\hat{e}^{(2)\mu}\hat{e}^{(2) \nu}\nonumber\\
&&+\left(f_3+f_4\slashed{P}\right)\left[\hat{e}^{(1) \mu}\hat{e}^{(2) \nu}-\hat{e}^{(2) \mu}\hat{e}^{(1) \nu}\right]~.
\label{amplitude2}
\eea
The first two terms vanish identically after inserting $F_{D}^{\mu\nu}$ in (\ref{amplitude1}). Moreover, one can show that the last term generates B-mode polarization for the CMB radiation \cite{zarei}.

We now turn to the case (ii) in which the amplitude is even under parity. The CPT invariance implies C-even and T-even or C-odd and T-odd. Among them, we are interested in the first condition plus crossing symmetry. Imposing these conditions yields \cite{zarei}
\bea
F_{D}^{\mu\nu}&&=\left(f_1+f_2\slashed{P}\right)\hat{e}^{(1)\mu}\hat{e}^{(1)\nu}+\left(f_3+f_4\slashed{P}\right)\hat{e}^{(2)\mu}\hat{e}^{(2) \nu}\nonumber\\
&&+f_5\gamma^{5}\slashed{P}\,\left[\hat{e}^{(1) \mu}\hat{e}^{(2) \nu}+\hat{e}^{(2) \mu}\hat{e}^{(1) \nu}\right]~,
\label{amplitude3}
\eea
that leads to the generation of circular polarization \cite{zarei}.

\subsection{Majorana neutrino}
It is also interesting to assume that the neutrinos are Majorana particles.
A Majorana fermion is a particle that is its own antiparticle and hence it has no electric charge \cite{Mohapatra:2004,Giunt:2007,Akhmedov:2014kxa}.
The Majorana spinor is  defined as $\psi_M=\gamma^0 C \psi^{\ast}_M$  where $C$ is the charge conjugation operator. The properties of Majorana bilinears under parity, charge conjugation and time reversal transformations have been summarized in \cite{Mohapatra:2004,Giunt:2007,Akhmedov:2014kxa}. The Majorana condition implies $\psi_M=\psi^{c}_M$. As a result, for Majorana spinor under charge conjugation we have
\beq
C^{-1}\psi_M \,C=\psi_M~.
\eeq
Therefore, in general one can write
\beq
C^{-1}\left(\bar{\psi}_M A\, \psi_M \right)C=\bar{\psi}_M A\, \psi_M~,
\eeq
where for $A=\gamma^{\mu}$
\beq
\bar{\psi}_M \gamma^{\mu}\, \psi_M =0~.
\eeq
Moreover, one can show that the transformations of the other Majorana bilinears under $P$, $T$ and $C$ are the same as Dirac bilinears. It has been discussed in \cite{Latimer:2016kdg,zarei} that the Compton scattering amplitude of photons with Majorana neutrino is given by
\bea
T^{M}_{fi}&&=\bar{u}_{r'}(q')\epsilon^{s}_{\mu}\Big[F_{M}^{\mu\nu}(q',p';q,p) \nonumber\\
&&+CF_{M}^{\nu\mu}(-q,p';-q',p)\transpose C^{-1}\Big]\epsilon^{s'}_{\nu}u_{r}(q)~.
\eea
Now, if in general
\beq
C F_{M}^{\nu\mu}(-q,p';-q',p)\transpose C^{-1}=-F_{M}^{\mu\nu}(q',p';q,p)~,
\eeq
we find that $T^{M}_{fi}=0$ identically. However, if
\beq
C F_{M}^{\nu\mu}(-p,-q;-p',-q')\transpose C^{-1}=F_{M}^{\mu\nu}(p,q;p',q')~,
\eeq
the scattering amplitude becomes twice as large for Majorana fermion.
Being invariant under the CPT transformations, the C-even property implies that the amplitude must be even under PT. Therefore, only possibilities are either P-even and T-even or P-odd and T-odd. The latter case means that CP is violated.

First, we consider the condition that amplitude is even under both P and T. We also consider the crossing symmetry, under which
\beq
e^{(1) \lambda} \leftrightarrow e^{(1) \lambda}  ~~~~~~~~ e^{(2) \lambda} \leftrightarrow -e^{(2) \lambda}~.
\eeq
After imposing the C-even together with P-even and T-even conditions
one can constrain the amplitude as
\bea
F_{M}^{\nu\mu}&=&\left(f_1+f_2\slashed{P}\right)\hat{e}^{(1)\mu}\hat{e}^{(1)\nu}+\left(f_3+f_4\slashed{P}\right)\hat{e}^{(2)\mu}\hat{e}^{(2) \nu}\nonumber\\
&&+f_5\gamma^{5}\slashed{P}\left[\hat{e}^{(1) \mu}\hat{e}^{(2) \nu}+\hat{e}^{(2) \mu}\hat{e}^{(1) \nu}\right]~,
\label{amplitude5}
\eea
that we again derive the same coefficients as the standard Compton scattering \eqref{amplitude3}. Hence, the amplitude for Majorana fermion can be written in terms of Dirac fermion amplitude as
\beq
T^{M}_{fi}=2T^{D}_{fi}~.
\label{DvM}
\eeq
For further investigation, let us consider the case that the amplitude is odd-P and odd-T. For this case, we find
\bea
F_{M}^{\nu\mu}&=&\left(f_1+f_2\slashed{P}\right)\left[\hat{e}^{(1) \mu}\hat{e}^{(2) \nu}-\hat{e}^{(2) \mu}\hat{e}^{(1) \nu}\right]~,
\label{amplitude6}
\eea
which is in agreement with \eqref{amplitude2}. As a result, for this case one can write $T^{M}_{fi}=2T^{D}_{fi}$.

\section{One-loop effective photon-neutrino Hamiltonian in forward scattering \label{sec:hamiltonian}}
This section is devoted to present an explicit example of possible photon-neutrino forward scattering in the framework of the SM that  predicts the properties described above. We also demonstrate a robust photon detection scheme using an optical cavity. 

We consider a cavity wall made by a perfect conductor and impose the periodic boundary condition on the photon field. Hence, in a rectangular cavity of volume $V$ we expand the photon field as
\beq
A^{\mu}(x)=\sum_{\mathbf{p},s}\left(\frac{\hbar c^2}{2V\omega_{\mathbf{p}}}\right)^{\hspace{-1.5mm}1/2}\hspace{-3mm}\epsilon^{\mu}_s(\mathbf{ p})\Big[a_s(\mathbf{ p})e^{-i\mathbf{ p}\cdot \mathbf{ x}}+a^{\dag}_s(\mathbf{ p})e^{i\mathbf{ p}\cdot \mathbf{ x}}\Big],\label{qphoton}
\eeq
where $\epsilon^{\mu}_s(\mathbf{p})$
are the photon polarization
four-vectors, the index $s = 1, 2$ labels the physical transverse polarizations of the photon and $p$ are discrete energy-momentum
values $\omega_{\mathbf{ p}}=p^0=|\mathbf{ p}|$ and
$p^i=2\pi \hbar cn^i/L$
with $n_i=0,\pm 1,\pm2,\cdot\cdot\cdot$, and $\sum_{\mathbf{ p}}$ is the sum over all cavity modes.
The photon annihilation $a({\bf p})$ and creation $a^{\dagger}({\bf p})$ obey \cite{mandl}
\beq\label{operator1}
 [a^\dagger_{s'}(\mathbf{p}'),a_{s}(\mathbf{p})]= \,\delta_{s,s'}\delta_{\mathbf{p},\mathbf{p}'}~.
 \eeq
The free neutrino field $\psi_\nu$ is also expanded as
\begin{align}
\psi_\nu(x)=\hspace{-1.5mm}\int\hspace{-1.5mm}\frac{d^3q}{(2\pi)^3}\frac{1}{\sqrt{2q^0}}&\big[u_r(\mathbf{ q})\,b_r(\mathbf{ q})e^{-i\mathbf{ q}\cdot \mathbf{ x}} \nonumber\\
&\hspace{-2mm}+v_r(\mathbf{ q})\,d^{\dag}_r(\mathbf{ q})e^{+i\mathbf{ q}\cdot \mathbf{ x}}\big]~,
\label{psi-nu}
\end{align}
where $u_r$ and $v_r$ are Dirac spinors, $b_r\,(d_r)$ and $b^\dagger_r \,(d^\dagger_r )$ are creation and annihilation operators for neutrino (anti-neutrino). The creation and annihilation operators  $b^\dagger_{r'}$ and $b_{r} $ obey the following relation
\begin{equation}\label{operator1}
  \{b^\dagger_{r'}(\mathbf{q}'),b_{r}(\mathbf{q})\}=(2\pi)^3\,\delta_{r'\,r}\delta^3(\mathbf{q}'-\mathbf{q}),
\end{equation}
and also  the expectational value over the neutrino creation and annihilation operators is introduced
\bea
\big\langle b^\dag_{r'}(q')b_{r}(q)\big\rangle =(2\pi)^3\delta^3(\mathbf{q}-\mathbf{q'})\,\delta_{rr'}\,n_\nu(\mathbf{x},\mathbf{q}),\label{exptation}
\eea
and $n_\nu(\mathbf{x},\mathbf{q})$ represents the local spatial density of neutrinos in the momentum state $\mathbf{q}$.
\begin{figure}[t]
\includegraphics[width=\columnwidth]{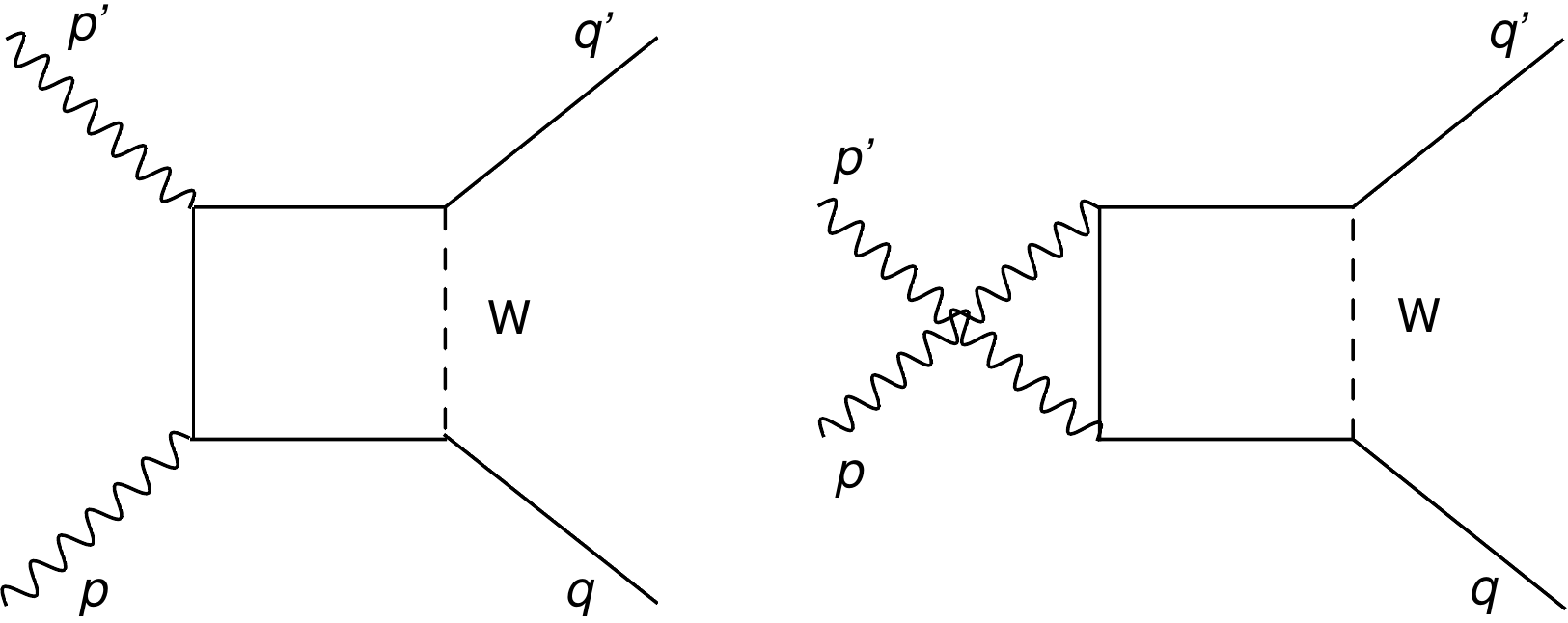}
\caption{Two possible Feynman diagrams for the photon-neutrino forward scattering amplitudes in the context of the SM.}\label{diagram}
\end{figure}

It has been shown in \cite{Mohammadi:2013ksa} that neutrinos can act as a birefringent background which leads to a small net circular polarization for a beam of photons, which undergo forward scattering with a beam of neutrino due to the interaction resulting from the one-loop correction of $W$ boson and charged lepton exchange (Fig.~\ref{diagram}). Moreover, it is predicted that the net circular polarization generated from the interaction with Majorana neutrinos doubles the net amount generated from the interaction with Dirac neutrinos. In following, we will try to reconstruct the mention results for the case that photons trapped in a cavity and interacting with neutrino beam. 
Note that the boundary condition would not be important as long as the time scale of photon-neutrino interaction $\tau_{\gamma \nu}$ is much smaller than $L/c$, and in this case, the trapped photons of the discrete spectrum can be approximated as free photons of continuous spectrum.

In the SM for elementary particle physics, the Hamiltonian of photon-neutrino forward scattering at the leading order is given by the exchange of leptons and the $W\text{-}Z$ gauge bosons, via the one-loop diagrams, as shown in Fig.~\ref{diagram},
\begin{widetext}
\bea
H_{\nu \gamma}&&=\!\int\! d\mathbf{ q}d\mathbf{ q}'\sum_{\mathbf{ p}}\Big(\frac{\hbar c^2}{2V\omega_{\mathbf{ p}}}\Big)^{\!1/2}\sum_{\mathbf{ p}'}\Big(\frac{\hbar c^2}{2V\omega_{\mathbf{ p'}}}\Big)^{\!1/2}
(2\pi)^3\delta^3(\mathbf{ q}'-\mathbf{ q})\delta_{\mathbf{ p},\mathbf{ p}'}\exp\l[i\left(q'^0+p'^0-q^0-p^0\right)t\r]
\nonumber \\
&&\times b^{\dagger}_{r'}(\mathbf{ q}')a^{\dagger}_{s'}(\mathbf{ p}')T_{fi}(ps,qr,p's',q'r')b_{r}(\mathbf{ q})a_{s}(\mathbf{ p}).
\label{h0}
\eea
The scattering amplitude $T_{fi}$ is given by
\begin{eqnarray}
T_{fi} &=& \frac{e^2g_w^2}{8}\int\frac{d^4l}{(2\pi)^4}D_{\alpha\beta}(q-l)\bar{u}_{r'}(\mathbf{q}')\gamma^\alpha (1-\gamma_5)\nonumber\\
&\times& S_F(l+p-p')\!\left[\ep_{s'}S_F(l+p)\ep_{s}+\ep_{s}S_F(l-p')\ep_{s'}\right]S_F(l) \gamma^\beta (1-\gamma_5)u_r(\mathbf{q})~,
\label{md}
\end{eqnarray}
where $\epsilon^{\mu}_s(\mathbf{p})$ and $\epsilon^{\mu}_{s'}(\mathbf{p}')$ are the transverse polarizations of initial and final states of cavity photons.
The wave functions $u_r(\mathbf{q})$ and $\bar{u}_{r'}(\mathbf{q}')$ are spinors of initial and final neutrinos, $r,r'$ label their spin degrees of freedom. The notations $D_{\alpha\beta}$ and $S_F$ represent the propagators of $W_\mu^\pm$ gauge-bosons and charged-lepton respectively.
The straightforward calculations of Eqs.~(\ref{h0}) and (\ref{md}) yield the one-loop effective Hamiltonian of the photon-neutrino forward scattering~\cite{Mohammadi:2013ksa}
\begin{eqnarray}
  H_{\nu \gamma} &=& \alpha\,G^F\frac{\sqrt{2}}{12\pi}\sum_{\mathbf{ p}}\left(\frac{\hbar c^2}{2V\omega_{\mathbf{ p}}}\right)a^{\dagger}_{s'}(\mathbf{p})\,a_s(\mathbf{p})\int \frac{d^3q}{(2\pi)^3}\,b^{\dagger}_{r}(\mathbf{q})\,b_r(\mathbf{q}) \nonumber \\
   &&\times\Big( 2\epsilon_{s'}(\mathbf{p})\cdot\epsilon_s(\mathbf{p})(q^0-|\mathbf{q}|)
  +4|{\bf q}|(\hat{\mathbf{q}}\cdot\epsilon_s)\,(\hat{\mathbf{q}}\cdot\epsilon_{s'})
  -2i\varepsilon_{\lambda\alpha\beta\mu}p^\alpha\,\epsilon_{s'}^\beta\,\epsilon_s^\mu\,\hat{q}^{\lambda} \Big),
\label{h2}
\end{eqnarray}
\end{widetext}
where $\hat q\equiv q/|q|$. We examine each term in 
Eq.~(\ref{h2}). The first term is zero for $s\not= s'$, however it gives a small contribution to the kinetic term $a^{\dagger}_{s}(\mathbf{p})\,a_s(\mathbf{p})$ for $s=s'$, as the mass-shell relation $q^0- |\mathbf{q}|\approx 0$ of neutrinos in the beam. This slightly modifies the frequency of cavity photons $\omega_{\bf p}\rightarrow \tilde\omega_{\bf p}$.  
The second term is proportional to neutrino energies and gives the dominate contribution to Eq.~(\ref{h2}). Instead, 
the third term is proportional to photon energies, 
and is about $\omega_{\mathbf{ p}}/|\mathbf{q}|$ times smaller than 
the second term. Thus the third term becomes negligible 
for the case $\omega_{\mathbf{ p}}/|\mathbf{q}|\ll 1$ of high-energy  neutrinos and low-energy microwave photons.
Next, by making the expectation value over the neutrino creation and annihilation operators (\ref{exptation}), 
\begin{equation}
\int \frac{d^3q}{(2\pi)^3}\,\left<b^{\dagger}_{r}(\mathbf{q})\,b_r(\mathbf{q})\right> = N_\nu~,
\label{number}
\end{equation}
where $N_\nu$ is the total number of neutrinos in the cavity during interacting time. As a result, we approximately obtain
\begin{equation}
H_{\nu \gamma} \simeq\sqrt{2}\! \alpha G^F\!\sum_{\mathbf{ p}}\frac{\hbar c^2 N_{\nu}|\mathbf{q}|}{6\pi\omega_{\mathbf{ p}}V}(\hat{\mathbf{q}}\cdot\epsilon_s)\,(\hat{\mathbf{q}}\cdot\epsilon_{s'})\,a^{\dagger}_{s}(\mathbf{p})\,a_{s'}(\mathbf{p})~,
\label{h4}
\end{equation}
where and henceforth $\mathbf{q}$ is considered as an averaged
energy-momentum of neutrino beam. The Hamiltonian (\ref{h4}) 
shows that photons possess the coupling and transition of two linear polarizations $s\not=s'$, thus they acquire the circular polarization $\Delta V^{D,M}$, 
when they interacts with neutrinos  ~\cite{Mohammadi:2013ksa}.    
By comparing Eq.~(\ref{h4}) with Eq.~\eqref{amplitude3}, one can easily set $f_2=f_3=f_4=f_5=0$ and find
\beq
T_{fi}(ps,qr,p's',q'r')\propto (\mathbf{\hat{q}}\cdot\bm{\epsilon}_s)(\mathbf{\hat{q}}\cdot\bm{\epsilon}_{s'})~.
\eeq
Therefore, we consider the case in which the amplitude $T_{fi}(ps,qr,p's',q'r')$ is C-even as well as P-odd and T-even.
 
We express the Hamiltonian (\ref{h4}) as an effective Hamiltonian
describing cavity photons interacting 
with neutrinos
\beq
H_{\nu \gamma}=\hbar  \sum_{\mathbf{ p},s,s'}\frac{ g_{\nu\gamma}}{\omega_{\mathbf{ p}}}(\mathbf{\hat{q}}\cdot\bm{\epsilon}_s)(\mathbf{\hat{q}}\cdot\bm{\epsilon}_{s'})\,a^{\dagger}_{s}a_{s'}~,
\label{Heff}
\eeq
where $g_{\nu\gamma}\equiv (\sqrt{2}/6\pi)\alpha G_F c \bar{F}_{\nu}$ is the coupling constant 
with $\bar{F}_{\nu}=N_\nu |\mathbf{ q}|c/V$ 
the energy flux of neutrino beam.
By substituting the values in SI units we arrive at
\beq
g_{\nu \gamma} \simeq 10^{-11} \bar{F}_\nu\Big[\frac{1}{{\rm GeV \cdot cm^{-2}\cdot s^{-1}}}\Big].
\label{gnugamma}
\eeq
The neutrino fluxes as large as $\bar{F}_\nu \approx 10^5~{\rm GeV\cdot cm^{-2}\cdot s^{-1}}$ are already available in T2K and Fermi-lab and one order of magnitude higher values are well within the reach.
Therefore, the problem of distinguishing Majorana from Dirac neutrino boils down to resolving processes with $g_{\nu\gamma}\simeq 10^{-5}$~Hz$^2$.
The effective Hamiltonian (\ref{h4}) refers to the case of a
Dirac neutrino, strictly speaking, a two-component
left-handed Weyl neutrino. Whereas, the effective
Hamiltonian for the case of Majorana fermions is
$H^{M}_{\nu \gamma}=2H^{D}_{\nu \gamma}=2H_{\nu \gamma}$.
The physical meaning is that the polarization flip due to Dirac neutrino scattering takes about $10^5$ seconds for a photon with frequency 1~Hz, while the process takes half of this time when scattered by Majorana neutrinos.
The reason is that the Majorana fermion
are  four-component and self-conjugated field
composed by a two-component left-handed Weyl field and its
conjugation and they both have the same contribution to
the effective Hamiltonian (\ref{h4}). As a result, the circular 
polarization amplitudes $\Delta V^{M}=2\Delta V^{D}$ in terms of Stokes parameter ~\cite{Mohammadi:2013ksa}.
This factor of two can make a possible
distinguish between Dirac and Majorana fermions by the
the experiment proposed below to measure the polarization $s$
transition of microwave photons trapped in an optical cavity and scattering off a Dirac or Majorana neutrino beam.

Before ending this section, we would like to point out that:
(i) in the forward scattering off neutrinos, the cavity photons retain their energy $\omega_{\bf p}$, but their polarization changes;
(ii) the interaction Hamiltonian (\ref{Heff}) represents the transition rate: per unit of time the number of photon polarization states flipping from one state $s$ to the other $s'$;
(iii) the lower the cavity photon frequency, the larger the scattering rate becomes.
It can be deduced from Eq.~(\ref{qphoton}) that for a given cavity volume $V$, the electromagnetic energy is proportional to $\omega^{-1}_{\bf p}$.
This justifies why the dominant contribution of the transition rate comes from the lowest lying cavity mode $\omega_{\bf p_0}=\pi c/L$, indicating the larger size of the cavity, the larger transition rate is.

\section{Detection scheme \label{sec:cavity}}
In this section, we propose an experiment for detecting the polarization conversion of the photons produced as a result of neutrino-photon interaction.
We show that thanks to the state-of-the-art ultrahigh finesse microwave cavities one can resolve the difference of Majorana and Dirac neutrinos through their scattering acceleration $g_{\gamma\nu}$.

The setup is a cavity placed next to a source of neutrino. We employ a doubly degenerate cavity mode supporting two orthogonal polarizations. We further assume that the cavity axis is aligned perpendicular to the propagation direction of neutrinos. This maximizes the scattering probability as Eq.~(\ref{Heff}) suggests.
The total Hamiltonian of the system is then
\bea
H/\hbar &=& \sum_\mathbf{p}\sum_{s=1,2}\tilde\omega_{\mathbf{p}} a_s^\dag(\mathbf{p}) a_s(\mathbf{p}) \nonumber\\
&+&\sum_\mathbf{p}J(\mathbf{p})\big[a_1^\dag(\mathbf{p}) a_2(\mathbf{p}) +a_1(\mathbf{p}) a_2^\dag(\mathbf{p}) \big]
\eea
where $\tilde\omega_\mathbf{p}$ is the modified cavity frequency in the presence of neutrinos and $J(\mathbf{p}) \equiv g_{\nu\gamma}/\tilde\omega(\mathbf{p})$ is the neutrino-induced coupling rate of the two orthogonal polarizations.
Here, we assume a mono-color drive with momentum $\mathbf{p}_0$. This then sets the active cavity mode, the mode that gets populated during the experiment. We assume that the intermode scattering rates are much smaller than the cavity free-spectral-range, therefore, the rest of the modes remain unpopulated. We also assume that the thermal excitation is negligible, which is justified for $\hbar\omega_\mathbf{p} \gg k_\text{B}T$. One thus simply works with a single cavity longitudinal mode and the simplified Hamiltonian in a frame rotating at the laser frequency $\Omega_{\mathbf{p}_0}$ reads
\beq
H/\hbar = \sum_{s=1,2}\Delta a_s^\dag a_s +J\big(a_1^\dag a_2 +a_1 a_2^\dag \big),
\label{Hsim}
\eeq
where we have dropped a $\mathbf{p}_0$ subscript for the sake of readability and introduced the laser detuning $\Delta \equiv \Delta_{\mathbf{p}_0} = \tilde\omega_{\mathbf{p}_0} - \Omega_{\mathbf{p}_0}$.
This Hamiltonian is quadratic, therefore, dynamics of the system is fully captured by the covariance matrix of the two mode quadratures.
The Langevin equations of motion for the annihilation operators are
\begin{subequations}
\begin{eqnarray}
\dot a_1 &=& -(\frac{\kappa_1}{2} +i\Delta)a_1 -i J a_2 +\sqrt{\kappa_1}\ a_1^{\rm in},\\
\dot a_2 &=& -(\frac{\kappa_2}{2} +i\Delta)a_2 -i J a_1 +\sqrt{\kappa_2}\ a_2^{\rm in},
\end{eqnarray}
\label{langevin}
\end{subequations}
where $\kappa_s$ with $s=1,2$ denotes loss rate of the cavity modes.
\begin{figure}[b]
\includegraphics[width=0.8\columnwidth]{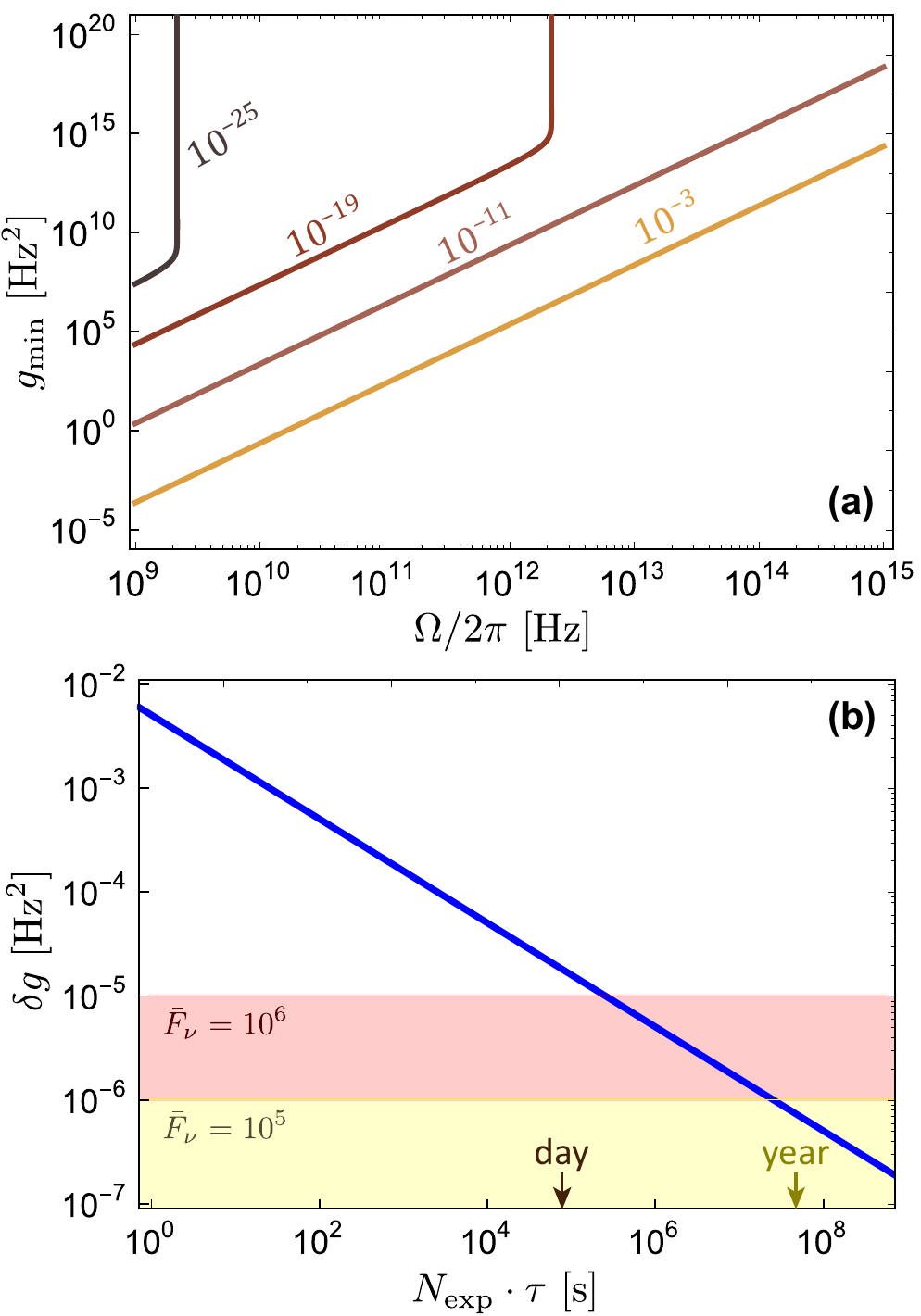}
\caption{
(a) The minimum resolvable neutrino scattering acceleration $g_{\rm min}$ in a single-shot experiment versus cavity photon frequencies at four different pump powers in units of Watts. Here, we have taken $Q=10^{10}$.
(b) Variations of the resolution $\delta g$ with respect to the total measurement time (blue solid line). The Majorana/Dirac resolution threshold is shown for two neutrino flux values $\bar{F}_\nu$: The state-of-the-art (yellow) and within reach in a near future (red) in units of ${\rm GeV\cdot cm^{-2}\cdot s^{-1}}$.
We set $P_1=1$~mW and $\Omega/2\pi=4.5$~GHz for the pump power and cavity mode frequency, respectively.
}
\label{fig:freq}
\end{figure}
In these equations $a_s^{\rm in}$ are the cavity input operators with the nonvanishing correlation functions
\begin{subequations}
\begin{align}
&\langle a_s^{\rm in}(t)a_{s}^{\rm in,\dag}(t')\rangle = (\bar{N}_s+1)\delta(t-t'), \\
&\langle a_s^{\rm in,\dag}(t)a_{s}^{\rm in}(t')\rangle = \bar{N}_s\delta(t-t'),
\end{align}
\end{subequations}
where $\bar{N}_s$ is composed of the thermal part and the coherent component. The thermal occupation number is given by $\bar{N}_{\rm th}(\tilde\omega) = \big[\exp\{\hbar\tilde\omega/k_{\rm B}T\} -1 \big]^{-1}$ at the temperature $T$, which we assume to be low enough to take $\bar{N}_{\rm th}\approx0$.
We assume that the laser input field only couples to the cavity mode $a_1$ and is related to the input power $P_1$ and normalized such that
\begin{equation}
\bar{N}_1\approx\frac{P_1}{\hbar\Omega}\cdot\tau_{\rm las},
\end{equation}
where $\tau$ is the average pulse duration and the small thermal part is neglected.
\begin{figure*}[t]
\includegraphics[width=0.8\textwidth]{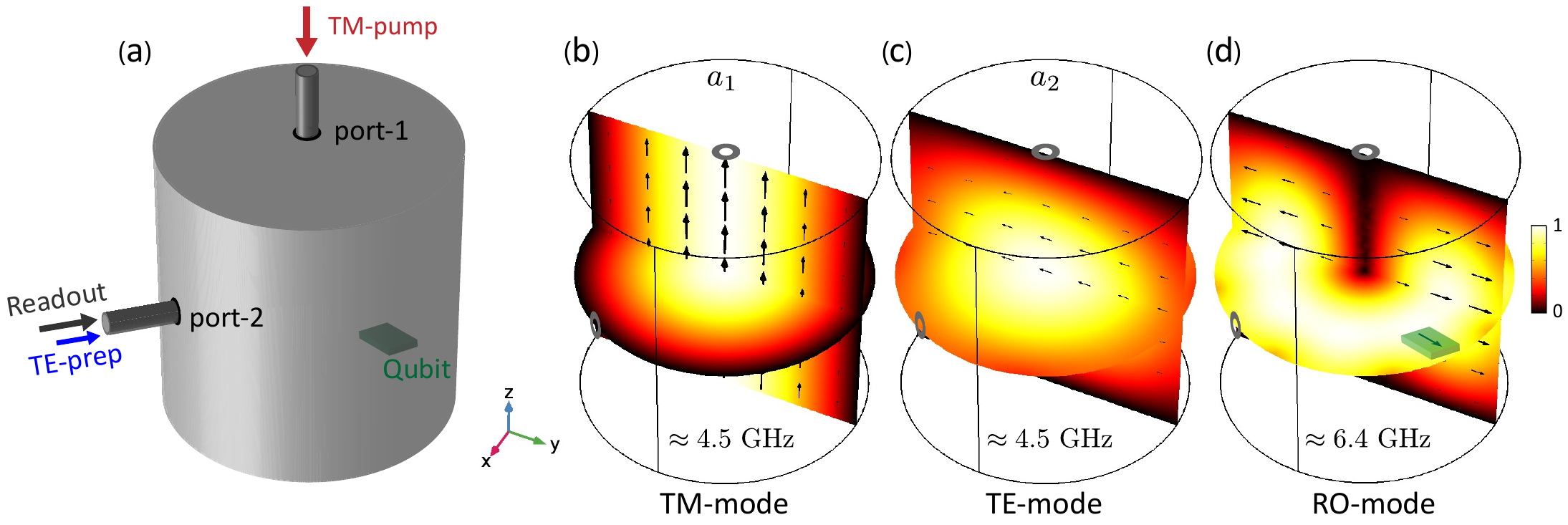}
\caption{Conceptual design of the experiment:
(a) The sketch of the proposed setup: a superconducting microwave cavity with two ports and a transmon qubit (the green slab) inside.
FEM simulation of the operative cavity modes: (b) The fundamental TM-mode used for pumping, (c) the symmetric degenerate TE-mode that accepts the neutrino-induced scattered photons, and (d) the readout mode which is TE in nature with a higher frequency. The transmon qubit is placed and oriented such that its coupling to the `pump' TE-mode is negligible due to the perpendicularity of its dipole moment (the green arrow) to the electric field vector of TM-modes.
The place of ports are shown as gray circles. The color bar shows the normalized electric field.
}
\label{fig:cavity}
\end{figure*}
In terms of the Hermitian quadrature operators introduced via $a_s = (x_s +i y_s)/\sqrt{2}$ the Langevin equations in (\ref{langevin}) take the form $\dot{\mathbf{u}}=\mathbf{A}\mathbf{u}+\mathbf{n}$ with $\mathbf{u}=[x_1,y_1,x_2,y_2]\transpose$ and $\mathbf{n}=[\sqrt{\kappa_1}x_1^{\rm in},\sqrt{\kappa_1}y_1^{\rm in},\sqrt{\kappa_2}x_2^{\rm in},\sqrt{\kappa_2}y_2^{\rm in}]\transpose$ and $\mathbf{A}$, the drift matrix, is given by
\beq
\mathbf{A} = \left[
\begin{array}{cccc}
	-\kappa_1/2 & \Delta & 0 & J \\
	-\Delta & -\kappa_1/2 & -J & 0 \\
	0 & J & -\kappa_2/2 & \Delta \\
	-J & 0 & -\Delta & -\kappa_2/2
\end{array}\right].
\eeq
The linearity of the Langevin equations of motion and Gaussian nature of the noise operators allows us to work with covariance matrix whose dynamics describes dynamics of the system through the following equation~\cite{Weedbrook2012}
\beq
\dot{\mathbf{V}} = \mathbf{A}\mathbf{V}+\mathbf{V}\mathbf{A}\transpose +\mathbf{D},
\label{lyapunov}
\eeq
where the diffusion matrix has been introduced as $\mathbf{D} = \frac{1}{2}\text{diag}[(2\bar{N}_1+1)\kappa_1,(2\bar{N}_1+1)\kappa_1,\kappa_2,\kappa_2]$ with assuming negligibility of the thermal photons.

We investigate the steady state case where duration of the laser drive pulse is longer than the cavity mode decay rates ($\tau_{\rm las}^{-1} \lesssim \min\{\kappa_1, \kappa_2\}$) and thus drags the system into a stationary state.
In this situation one sets the left-hand-side in Eq.~(\ref{lyapunov}) equal to  zero and solves for time-independent $\mathbf{V}_{\rm ss}$.
The expectation value of the number operator $\langle a_2^\dag a^{}_2\rangle_{\rm ss}$ in the steady-state is thus analytically obtained. This quantity determines the number of the neutrino-induced scattered photons.
The minimum coupling acceleration $g_{\nu\gamma}^{\rm min}$ that leads to the scattering of at least one photon from $a_1$ into $a_2$ at resonance ($\Delta=0$) and taking $\tau_{\rm las} \simeq 1/\kappa_1$ is
\begin{equation}
g_{\nu\gamma}^{\rm min} = \frac{\Omega}{2}\Big[\frac{\kappa_1\kappa_2(\kappa_1 +\kappa_2)}{\frac{P_1}{\hbar\Omega} -(\kappa_1 +\kappa_2)}\Big]^{\frac{1}{2}}.
\label{gmin}
\end{equation}
There is a linear relation between the cavity frequency and the minimum detectable $g_{\nu\gamma}^{\rm min}$. Hence, working with the low-frequency photons is more efficient. Without loss of generality we assume equal values for the two cavity decay rates $\kappa_1 =\kappa_2$ in the rest of article.
This simplifies Eq.~\eqref{gmin} and one easily notices that as expected by increasing the input laser power the number of cavity photons increases and hence the probability of scattering. Therefore, for $N_1 \gg 1$ we get $g_{\nu\gamma}^{\rm min} \approx (\Omega^2/2)\sqrt{2\hbar/P_1Q}$ where $Q$ is the quality factor of the two cavity modes.
For several input powers the single-shot resolution $g_{\nu\gamma}^{\rm min}$ is plotted in Fig.~\ref{fig:freq}(a).
Note that as discussed below Eq.~\eqref{gnugamma}, the scattering rate can be measured once the polarization flip is observed. Nevertheless, the limit at which the difference of Dirac and Majorana neutrinos becomes detectable is set at about $g_{\nu\gamma}^{\rm min} \simeq 10^{-6}$~Hz$^2$ with the current technology (see below), where the number of scattered photons differs by a factor of two.
Therefore, the measurement precision needs to be enhanced by e.g. repeating the experiment.

The classical ergodicity implies that by repeating the experiment for $N_{\rm exp}$ times the final achievable resolution for the photon-neutrino scattering acceleration improves by $\delta g \equiv g_{\nu\gamma}^{\rm min}/\sqrt{N_{\rm exp}}$. Therefore, the total runtime for detecting/excluding neutrino-induced scattered photons is then $t = N_{\rm exp}\tau$, where $\tau$ is the time period composed of the pump pulse and the readout process. In this work we consider longtime pump pulses, therefore, 
in Fig.~\ref{fig:freq}(b) the achievable resolution is plotted against total experiment duration. The results suggest that the neutrino-induced photon--photon scattering in the level of distinguishing between Majorana and Dirac neutrinos is detectable/excludable in a few days when the neutrino flux is $\bar{F}_\nu=10^6~\text{GeVcm}^{-2}\text{s}^{-1}$. The process is, nonetheless, anticipated to take about a year with the currently available neutrino flux values.
In the following we provide details of the proposed experiment.


\section{Implementation \label{sec:implement}}
To implement the scheme explained above we propose to use a superconducting microwave cavity as they benefit lowest attainable frequency without appreciable thermalization at dilution refrigerator temperatures and yet are among those with highest quality factors~\cite{Kuhr2007, Romanenko2014}. Two properly chosen degenerate modes; one TE and the other TM of frequency $\Omega/2\pi = 4.5$~GHz are used for the scattering experiment, while a third ancillary mode with frequency $\Omega_{\rm ro}/2\pi = 6.4$~GHz is employed for readout (we shall call it the RO mode).
This can be realized, for example, in a cylindrical cavity of length $L\approx 5.2$~cm and radius $R\approx 2.6$~cm [see Fig.~\ref{fig:cavity} for the sketch and finite element simulation].
Two distinct in/out ports are appropriately positioned on the cavity walls. The first one (port-1) is used to pump the fundamental TM-mode of the cavity (the $a_1$ mode in the above analyses), while the other one is strongly coupled to the readout cavity mode.
To detect the photonic state of the $a_2$ TE-mode---which with a finite probability is expected to be a single-photon as a result of neutrino induced conversion of the pump photons---we propose to position and orient a transmon superconducting qubit inside the cavity such that it only couples to the TE and RO modes~\cite{Gasparinetti2016, Brecht2017, Xie2018}. The mode profiles and orientation of the electric field makes this possible as confirmed by the finite-element simulations.
Transmon qubits are known for their long coherence times and thus are proper candidates for the readout part in our protocol. The qubit frequency is chosen to be in resonance with the TM cavity mode ($\omega_q=\Omega$) and in strong dispersive coupling to the RO mode~\cite{Schuster2007}. The former will result-in coherent population transfer from the TE-mode state to the transmon qubit and the latter allows for rapid high-fidelity single-shot readout of the qubit~\cite{Walter2017}.

The detection scheme is composed of the following pulse sequence: (i) A preparation pulse sent through port-2 excites the TE-mode and resets the transmon qubit. (ii) Subsequently, a strong microwave pump resonantly drives the TM-mode via port-1. (iii) After that the state of transmon qubit is readout through a single-shot dispersive scheme by a pulse with central frequency $\Omega_{\rm ro}$ sent into port-2.
Since the qubit only can get excited by the TE-mode detecting an excited qubit signals scattering of a photon into the TE-mode.
\begin{figure}[b]
\includegraphics[width=0.8\columnwidth]{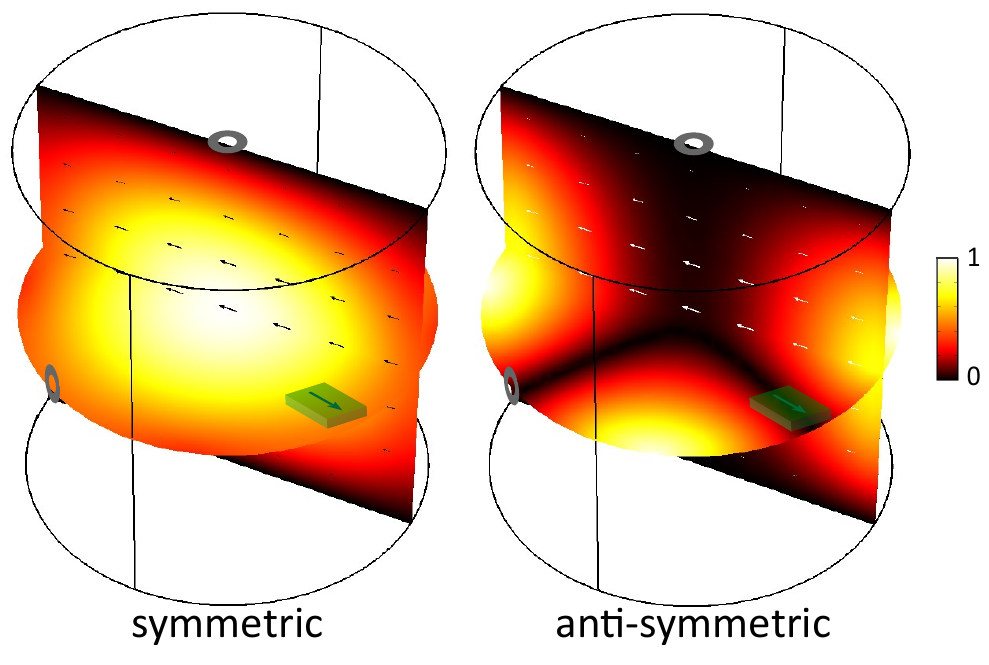}
\caption{%
FEM simulation of the electric field profile of the symmetric and anti-symmetric TE-modes which are degenerate with the fundamental TM-mode. The small overlap of the anti-symmetric mode with the TM-mode is much smaller than the symmetric one. The position of qubit (green slab) is chosen to only couple to the symmetric TE-mode (see the text for a detailed discussion).
}
\label{fig:anti}
\end{figure}

\subsection{Finite-element simulation}
To present a conceptual design for our scheme, here we provide a simulation based on finite-element method (FEM).
The results are summarized in Fig.~\ref{fig:cavity} alongside a sketch of the cavity and the ports.
The cavity is taken cylindrical whose dimensions are: The radius $2.56$~cm and the height $5.16$~cm. The fundamental mode is transverse magnetic (TM) with a Gaussian profile around the cavity axis (the axis of cylinder) such that the electric field is vanishing on the side wall. With the chosen dimensions frequency of this mode becomes $\Omega/2\pi\approx 4.5$~GHz.
The transmon qubit is oriented such that its dipole moment is perpendicular to the electric field vectors of the transverse magnetic modes. Therefore, its coupling to the TM-mode is negligible.
At the same frequency one finds two transverse electric (TE) modes: one symmetric and the other anti-symmetric.
We propose to employ the symmetric mode as a host for the scattered photons.
The profile suggests that overlap of the anti-symmetric mode with the TM-mode is much smaller than the symmetric one [compare the mode profiles in Fig.~\ref{fig:anti}]. We, therefore, expect negligible scattering into the anti-symmetric mode and ignore its effect in our considerations. This reasoning is further supported by positioning the transmon qubit in one of its nodes.
A higher transverse electric mode with frequency $\Omega_{\rm ro}/2\pi \approx 6.4$~GHz is proposed to employ for the readout process. The mode is TE and has an anti-node at the position of the qubit, hence, couples to it in a dispersive strong fashion. Moreover, it has an anti-node at the readout port, which guarantees a fast evacuation, and thus, rapid identification of the qubits state~\cite{Walter2017}.

\section{Conclusion and remarks \label{sec:conclusion}}
To summarize, we have studied the photon-neutrino scattering amplitude for both Dirac and Majorana neutrino cases. A one-loop effective interaction Hamiltonian then has been derived for forward scattering of photon polarization. We find that the neutrino-induced photon-photon polarization scattering rate differ in a factor of two. That is, the Majorana neutrinos should scatter polarization of photons twice larger than the Dirac neutrinos in the same time duration.
In order to resolve this difference, we have proposed an experimental setup that photons are scattered inside a cavity. The results suggest that the experiment performs much better for low frequency photons.
We, therefore, have designed and simulated a microwave cavity where one of its modes ($a_1$ TM-mode) is employed for pumping photons, one other mode with a perpendicular polarization ($a_2$ TE-mode) accommodates the scattered photons. The scattered photons are then absorbed by a superconducting qubit (a transmon qubit). The state of qubit, in turn, is quickly readout by and auxiliary off-resonance cavity mode (RO-mode).
The high quality factors of the current technology microwave and radio-frequency cavities allow one to probe the difference in the course of a year. However, we anticipate that by near future technologies the experiment timing should be reduced down to a few days. An enhancement in the neutrino beam flux yet reduces it to a few hours.

Before ending this article, we would like to mention that the theoretical and experimental studies presented in this article can be possibly generalized to the fields of studying the right-handed neutrinos, their possible small coupling to gauge bosons and 
flavor mixing, as well as 
other issues of probing new physics in the lepton sector, 
see for example the review article \cite{review_A} and Refs.~\cite{Xue2016}. Based on recently rapid developed technologies of lasers and microwave, such an experiment presented in this article is indeed a complementary approach to study the lepton-sector physics in the SM, in addition to the traditional experiments in nuclear and particle physics.

\begin{acknowledgments}
The authors thank S. Matarrese, A. Bettini, and G. Carugno for useful discussions. SSX thanks M. Dietrich for discussions and communications. MZ thanks  Department of Physics and Astronomy “G. Galilei” at University of Padova and ICRANet at Pescara for the hospitality while this work was in progress.
MA acknowledges support by STDPO and IUT through SBNHPCC.
\end{acknowledgments}


\end{document}